\documentclass[
reprint,
superscriptaddress,
amsmath,amssymb,
aps,pra,
floatfix,
]{revtex4-2}

\usepackage{float}
\usepackage[utf8]{inputenc}
\usepackage{graphicx,import}
\graphicspath{{Figures/}}
\usepackage{bm}
\usepackage[colorlinks]{hyperref}
\usepackage[all]{hypcap}
\usepackage{xcolor}
\usepackage{xfrac}
\usepackage{braket}
\usepackage{siunitx}
\usepackage{changes}
\usepackage{layouts}
\usepackage{anyfontsize}

\definecolor{bluegray}{RGB}{40,180,160}
\definecolor{navygray}{RGB}{110,140,170}
\definecolor{meadowgreen}{RGB}{0,128,0}
\definecolor{coolbrown}{RGB} {165,42,42}

\DeclareSIUnit{\sq}{\Box}

\newcommand{\secref}[1]{\hyperref[#1]{{Section~\ref{#1}}}}
\newcommand{\chapref}[1]{\hyperref[#1]{{Chapter~\ref{#1}}}}
\newcommand{\suppref}[1]{\hyperref[#1]{{App.~\ref{#1}}}}
\newcommand{\figref}[1]{\hyperref[#1]{{Fig.~\ref*{#1}}}}
\newcommand{\Figref}[1]{\hyperref[#1]{{Figure~\ref*{#1}}}}
\newcommand{\figrefadd}[2]{\hyperref[#1]{{Fig.~\ref*{#1}#2}}}
\newcommand{\Figrefadd}[2]{\hyperref[#1]{{Figure~\ref*{#1}#2}}}
\newcommand{\tabref}[1]{\hyperref[#1]{Tab.~\ref*{#1}}}
\newcommand{\refref}[1]{\hyperref[#1]{{Ref.~\ref*{#1}}}}
\renewcommand{\eqref}[1]{\hyperref[#1]{{Eq.~(\ref*{#1})}}}

\newcommand{\revisex}[1]{{}}

\hypersetup{
citecolor={navygray}, 
linkcolor={navygray},
urlcolor={navygray},
}

\makeatletter
\newcommand*{\balancecolsandclearpage}{%
  \close@column@grid
  \cleardoublepage
  \twocolumngrid
}
\makeatother

\begin{document}
\title{
Thermalization of a flexible microwave stripline measured by a superconducting qubit}

\author{Patrick~Paluch}
\email{patrick.paluch@kit.edu}
\affiliation{IQMT,~Karlsruhe~Institute~of~Technology,~76131~Karlsruhe,~Germany}
\affiliation{PHI,~Karlsruhe~Institute~of~Technology,~76131~Karlsruhe,~Germany}

\author{Martin~Spiecker}
\affiliation{IQMT,~Karlsruhe~Institute~of~Technology,~76131~Karlsruhe,~Germany}
\affiliation{PHI,~Karlsruhe~Institute~of~Technology,~76131~Karlsruhe,~Germany}

\author{Nicolas~Gosling}
\affiliation{IQMT,~Karlsruhe~Institute~of~Technology,~76131~Karlsruhe,~Germany}
\affiliation{PHI,~Karlsruhe~Institute~of~Technology,~76131~Karlsruhe,~Germany}

\author{Viktor~Adam}
\affiliation{IQMT,~Karlsruhe~Institute~of~Technology,~76131~Karlsruhe,~Germany}
\affiliation{PHI,~Karlsruhe~Institute~of~Technology,~76131~Karlsruhe,~Germany}

\author{Jakob~Kammhuber}
\affiliation{Delft Circuits B.V., Lorentzweg 1, 2628CJ Delft, The Netherlands}

\author{Kiefer~Vermeulen}
\affiliation{Delft Circuits B.V., Lorentzweg 1, 2628CJ Delft, The Netherlands}

\author{Dani\"{e}l~Bouman}
\affiliation{Delft Circuits B.V., Lorentzweg 1, 2628CJ Delft, The Netherlands}

\author{Wolfgang~Wernsdorfer}
\affiliation{IQMT,~Karlsruhe~Institute~of~Technology,~76131~Karlsruhe,~Germany}
\affiliation{PHI,~Karlsruhe~Institute~of~Technology,~76131~Karlsruhe,~Germany}

\author{Ioan~M.~Pop}
\email{ioan.pop@kit.edu}
\affiliation{IQMT,~Karlsruhe~Institute~of~Technology,~76131~Karlsruhe,~Germany}
\affiliation{PHI,~Karlsruhe~Institute~of~Technology,~76131~Karlsruhe,~Germany}
\affiliation{PI1,~Stuttgart~University,~70569~Stuttgart,~Germany}

\date{\today}

\begin{abstract}
With the demand for scalable cryogenic microwave circuitry continuously rising, recently developed flexible microwave striplines offer the tantalyzing perspective of increasing the cabling density by an order of magnitude without thermally overloading the cryostat. 
We use a superconducting quantum circuit to test the thermalization of input flex cables with integrated 60~dB of attenuation distributed at various temperature stages. 
From the measured decoherence rate of a superconducting fluxonium qubit, we estimate a residual population of the readout resonator below $3.5\cdot10^{-3}$ photons and we measure a $0.28\,$ms thermalization time for the flexible stripline attenuators.
Furthermore, we confirm that the qubit reaches an effective temperature of $26.4\,\mathrm{mK}$, close to the base temperature of the cryostat, practically the same as when using a conventional semi-rigid coaxial cable setup.
\end{abstract}

\maketitle
\section{Introduction}
The growing size of cryogenic quantum processors~\cite{Google2023Feb, Dupont2023Nov, Kim2023Jun, Krinner2022May, Zhu2022Feb} and detector arrays~\cite{Wollman2019Nov, Catalano2018Dec, Gottardi2021Apr, Gastaldo2017Jun} requires an increasing microwave circuitry density for readout and control. Possible strategies to cope with this challenge consist in frequency or time-division multiplexing~\cite{Kempf2017Jan, Dober2021Feb, Durkin2021Mar, Chen2012Oct, Heinsoo2018Sep}, which are, however, limited by the available bandwidth and the finite lifetime of the measured states. This motivates the demand for increasingly denser cryogenic microwave circuitry compatible with high-coherence devices~\cite{Krinner2019Dec}, for which new platforms based on photonic links~\cite{Lecocq2021Mar, Youssefi2021May, Shen2024Apr} or flexible microwave striplines~\cite{Pappas2016Jul, Tuckerman2016Jul, Walter2017Nov, Gupta2019Mar, Zou2019May, Smith2020Jul, Monarkha2024May} have recently been developed.

Flexible striplines with integrated microwave attenuators and filters promise to increase the cabling density by at least one order of magnitude compared to conventional coaxial setups.
Here we use a superconducting fluxonium qubit in a circuit quantum electrodynamics (QED) readout architecture to measure their in-situ thermalization time and contribution to photon shot noise dephasing.
We show that the flexible stripline thermalizes with a time constant of $0.28\,$ms, almost a factor of two faster than cryogenic coaxial attenuators in a similar setup, and we measure residual photon populations below $3.5\cdot10^{-3}$.
These findings, combined with the fact that we do not observe any detrimental effects on the superconducting device, recommend the use of flexible striplines at scale in future quantum processor setups and large detector arrays.

\section{Measurement concept and setup}
\begin{figure*}[t!]
\centering
\includegraphics[width=0.94\textwidth]{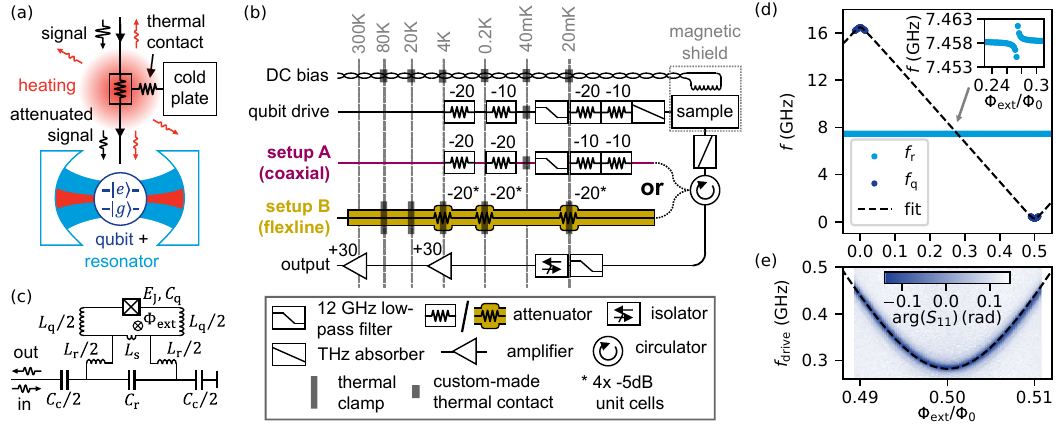}
\caption{
Cryogenic microwave setup and resonator-qubit device.
(a)~Principle of measuring the thermalization of the input lines. Microwave power dissipated in the attenuators generates local heating which radiates towards the resonator-qubit device, deteriorating its performance.
(b)~Microwave setup with a direct current (DC) line for magnetic flux biasing of the fluxonium, and three microwave lines for qubit control, readout signal input and output. The input line is implemented either as a coaxial cable with \textit{SubMiniature A} (SMA) connectorized attenuators (setup A) or a flexible coplanar stripline with integrated attenuators (model \textit{Cri/oFlex}® 3, ~\cite{DelftCircuits}), consisting of $5\,$dB unit cells (setup B). Both readout assemblies contain a total of $60\,$dB attenuation, distributed at various temperature stages. In setup A, the attenuators at $20\,$mK were additionally thermalized via a copper braid (\suppref{sec:supp:readoutlines_coax}). The coaxial lines are thermalized to the $40\,$mK intermediate stage using Ag-plated Cu wires and contain a 12$\,$GHz low-pass filter anchored at $20\,$mK. In setup B, we use custom-designed thermal clamps at $80\,$K, $20\,$K, $4\,$K, $200\,$mK and $20\,$mK, as detailed in \suppref{sec:supp:thermalizationexperiments} and \ref{sec:supp:readoutlines_flex}. In the output line, signals are amplified by $60\,$dB using a high-electron-mobility transistor (HEMT) at $4\,$K and a room-temperature amplifier. The sample is surrounded by an aluminum and mu-metal shield, as used in Ref.~\cite{Grunhaupt2018Sep}.
(c)~Simplified electrical circuit diagram of the fluxonium with inductance $L_\mathrm{q}+L_\mathrm{s}$, capacitance $C_\mathrm{q}$ and Josephson energy $E_\mathrm{J}$ and external flux bias $\Phi_\mathrm{ext}$. The qubit is coupled inductively to a readout resonator with fundamental mode frequency $f_\mathrm{r}$. The readout resonator is coupled via the capacitance $C_\mathrm{c}$ to the microwave lines.
(d)~Measured resonator and qubit frequencies $f_\mathrm{r, q}$ as a function of the external flux $\Phi_\mathrm{ext}$. From the fit (dashed line), we extract the qubit parameters (\tabref{tab:params}). The inset highlights the avoided level crossing between the qubit and resonator modes.
(e)~Two tone spectroscopy of the qubit around $\Phi_\mathrm{ext} / \Phi_0 \approx 0.5$.
}
\label{fig:sample}
\end{figure*}
In the last two decades, superconducting qubits emerged as one of the most promising candidates for future large-scale quantum processors~\cite{Google2023Feb, Dupont2023Nov, Kim2023Jun, Krinner2022May, Zhu2022Feb}.
One reason for this development is the steadily increasing qubit coherence time, nowadays exceeding a few hundred microseconds~\cite{Kjaergaard2020Mar, Somoroff2023Jun, Bal2024Apr}. 
This improvement also results in a higher sensitivity to dephasing, a measure for fluctuations of the qubit frequency, that originates from a multitude of different noise sources~\cite{Zimmerli1992Jul, Wellstood1987Mar, VanHarlingen2004Aug}.
The standard tool to read out quantum information in circuit QED is the dispersive coupling of the qubit to a readout resonator or cavity~\cite{Blais2021May}, which, however, adds another noise source for dephasing.
This so-called photon shot noise arises from the fact that each photon in the resonator changes the qubit frequency by the dispersive shift $\chi$~\cite{Schuster2005Mar, Gambetta2006Oct, Sears2012Nov}.
In this way, fluctuations in the average resonator photon number $\bar{n}$ directly translate into qubit dephasing.

Excess photons in the resonator originate from heat loaded in attenuators or filters anchored at higher temperature stages, transmitted via the microwave lines in form of black-body radiation~\cite{Yeh2017Jun} (\figrefadd{fig:sample}{(a)}).
In experiments with resonators in the gigahertz (GHz) regime, the residual $\bar{n}$ is observed to be between $2\cdot10^{-4}$ and $2\cdot10^{-1}$~\cite{Wang2019Jan, Yeh2017Jun, Yan2016Nov, Yan2018Jun, Suri2013Dec, Rigetti2012Sep, Goetz2017Mar, Zhang2017Jan, Rieger2023Feb}, orders of magnitude larger than the expected $\sim10^{-8}$ when in thermal equilibrium at $20\,$mK.
While the lower observed limit corresponds to dephasing rates that are on the edge of measurability with state-of-the-art superconducting qubits~\cite{Wang2019Jan}, coherence times are usually dominated by photon shot noise in the upper limit.
Therefore, it is crucial to ensure that new microwave input lines do not degrade qubit performance by causing excessive photon shot noise-induced dephasing.

\begin{table}[b]
\begin{tabular}{c|c|c|c|c|c}
\centering
$f_\mathrm{r}\,$(GHz) & $\sfrac{\kappa}{2\pi}\,$(MHz)  & $\sfrac{\chi}{2\pi}\,$(MHz) & $L_\mathrm{q}\,$(nH) & $C_\mathrm{q}\,$(fF) & $E_\mathrm{J}\,$(GHz) \\ \hline
7.458 & 4.10 & -2.70 & 176 & 5.73 & 16.6
\end{tabular}
\caption{Parameters of the resonator-qubit device.}
\label{tab:params}
\end{table}
Here, we exploit photon shot noise to quantify the thermalization of different microwave input lines.
Besides the passive heat load from higher temperature stages, the attenuators in the input line are also heated by readout and control pulses.
This active heat load generates additional photon shot noise-induced dephasing~\cite{Yeh2017Jun, Wang2019Jan}, which we use to quantify the thermal contact between the attenuator and the cold plate.
In separate cooldowns of our dilution refrigerator (model \textit{Sionludi XL},~\cite{qinu}), we compare the performance of two microwave input lines, a conventional coaxial cable and a flexible stripline (\figrefadd{fig:sample}{(b)}).
The microwave setup following the input lines remains unchanged in both cases.
Furthermore, we use a separate drive line for qubit manipulation and a DC line for magnetic flux biasing.
\begin{figure*}[t!]
\centering
\includegraphics[width=\textwidth]{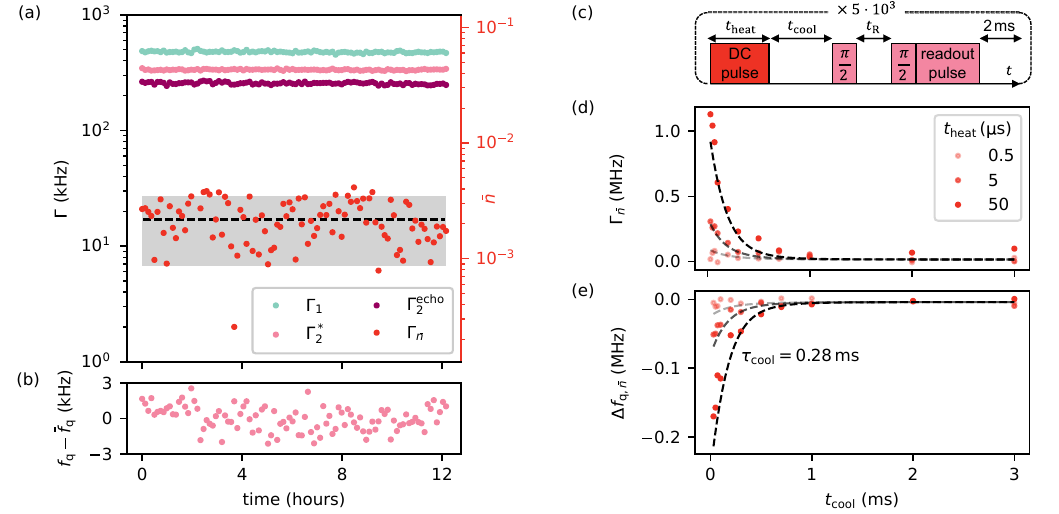}
\caption{
Fluxonium qubit measurements using the flexible stripline readout at $\Phi_\mathrm{ext}/\Phi_0 = 0.5$.
(a)
Interleaved measurements of $\Gamma_1$ relaxation (turquoise), $\Gamma_2$ Ramsey (pink) and echo (magenta) dephasing rates over 12 hours. Gaussian distribution fits yield values of $\Gamma_1 = (477\pm9)\,$kHz, $\Gamma_2^* = (335\pm4)\,$kHz and $\Gamma_2^\mathrm{echo} = (256\pm6)\,$kHz, respectively. Photon shot noise-induced dephasing rates $\Gamma_{\bar{n}}$ (red) are calculated from the extracted $\Gamma_1$ and $\Gamma_2^\mathrm{echo}$ values as described in the main text and converted into resonator photon numbers $\bar{n}$ according to \eqref{eq:eq2}. 
(b)
Qubit frequency extracted from Ramsey fringes yield average fluctuations $\sqrt{\langle{(f_\mathrm{q}-\bar{f}_\mathrm{q})^2}\rangle} = 1.0\, \mathrm{kHz}$ around the average qubit frequency $\bar{f}_\mathrm{q}$.
(c)
Pulse sequence for the measurement of attenuator thermalization time. We send a DC pulse of duration $t_\mathrm{heat}$ in the readout line to heat the attenuator, followed by a variable wait time $t_\mathrm{cool}$ before a standard Ramsey sequence.
(d)
Photon shot noise-induced dephasing rates $\Gamma_{\bar{n}}$ as well as 
(e)
qubit frequency shifts $\Delta f_{\mathrm{q},\bar{n}}$ extracted from Ramsey measurements. The dashed lines show an exponential fit with common temperature relaxation time $\tau_{\mathrm{cool}}=0.28\,\mathrm{ms}$ for all heat pulses, following \eqref{eq:dephasing} and \eqref{eq:heat-fit}.
}
\label{fig:meas1}
\end{figure*}

In order to assess information about photon shot noise, we use a superconducting quantum circuit consisting of a fluxonium qubit, inductively coupled to a resonator (\figrefadd{fig:sample}{(c)}), implementing the dispersive readout scheme. 
As described in \suppref{sec:supp:ChiKappaPower}, we extract the resonator bandwidth $\kappa / 2\pi$ and the dispersive shift $\chi / 2\pi$ from fits to the qubit state-dependent response of the readout resonator.
The qubit spectrum is depicted in \figrefadd{fig:sample}{(d)}, while a close-up around its first-order flux-insensitive point $\Phi_\mathrm{ext}/\Phi_0 = 0.5$ is shown in \figrefadd{fig:sample}{(e)}.
Here $\Phi_\mathrm{ext}$ denotes the external magnetic flux threading the fluxonium loop and $\Phi_0$ is the magnetic flux quantum.
Relevant parameters of the resonator-qubit system are summarized in \tabref{tab:params}.

\section{Results and discussion}
\begin{figure*}[t!]
\centering
\includegraphics[width=1.0\textwidth]{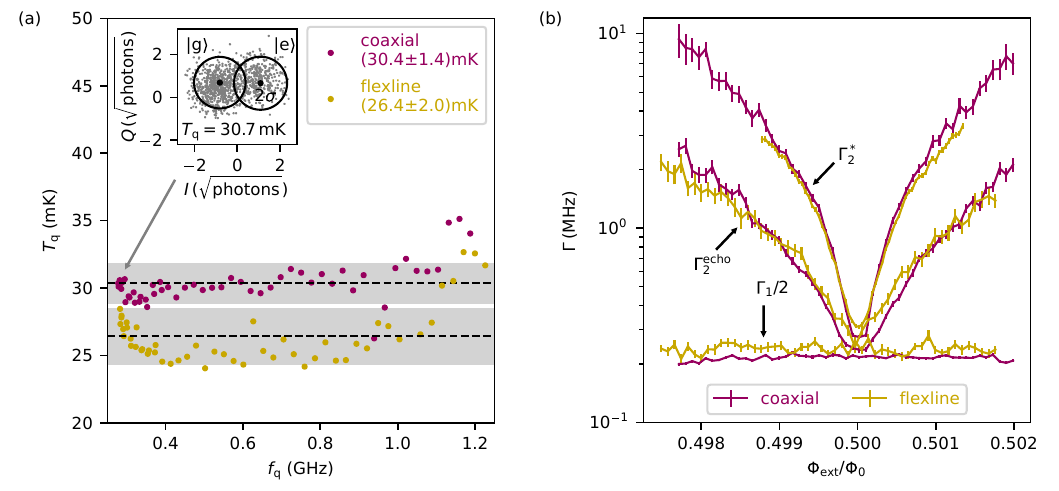}
\caption{
Comparison of conventional coaxial cable (magenta) and flexible stripline (yellow) setups.
(a)~Temperature $T_\mathrm{q}$ as a function of the qubit frequency $f_\mathrm{q}$, extracted from Gaussian mixture fits of measured $IQ$ distributions. We fit $10^5$ points for the coaxial cable and $5\cdot10^4$ for the flexible stripline. The dashed lines indicate the mean temperature values $\bar{T_\mathrm{q}}$ and the horizontally filled areas the $\pm1\sigma$ range, as stated in the legend. The inset shows a measured $IQ$ distribution using the coaxial line setup at $\Phi_\mathrm{ext} / \Phi_0 = 0.5$, as indicated by the arrow. For better visibility, only a subset of $2000$ points is shown. The quadratures are normalized to the square-root of the number of measurement photons $n_\mathrm{meas} \approx \bar{n} \kappa t_\mathrm{meas} / 4$, where $\bar{n}$ is the average number of photons in the resonator and $t_\mathrm{meas}$ the duration of the readout pulse. The black circles indicate the $2\sigma$ regions centered on the pointer states (black markers) corresponding to the qubit in the ground state $\vert g\rangle$ and the excited state $\vert e\rangle$, as indicated by the labels.
(b)~Rates (with vertical errorbars from the fit) extracted from energy relaxation, Ramsey as well as echo experiments in the vicinity of $\Phi_\mathrm{ext} / \Phi_0 = 0.5$, for both the coaxial cable and the flexible stripline setup.
}
\label{fig:meas2}
\end{figure*}
To infer the photon shot noise due to passive heat load in the flexline setup, we perform energy relaxation, Ramsey and echo measurements over a course of 12 hours in an interleaved manner (\figrefadd{fig:meas1}{(a)}).
The measured energy relaxation rates $\Gamma_1$ and decoherence rates $\Gamma_2^{*/\mathrm{echo}}$ fluctuate within 2\% and the qubit frequency, extracted from the Ramsey fringes, is stable within a few kHz (\figrefadd{fig:meas1}{(b)}).
We estimate photon shot noise induced dephasing rates $\Gamma_{\bar{n}} \approx \Gamma_\varphi = \Gamma_2 - \Gamma_1 / 2$ by using $\Gamma_2 = \Gamma_2^\mathrm{echo}$ to extract the fast components of the phase noise that can be associated with the photon shot noise. 
This results in values for $\Gamma_{\bar{n}}$ between 7$\,$kHz and 27$\,$kHz, that we convert into residual photon numbers $\bar{n}$ between $0.9\cdot10^{-3}$ and $3.5\cdot10^{-3}$, which is in the lower range of what is commonly observed in the community~\cite{Wang2019Jan, Yeh2017Jun, Yan2016Nov, Yan2018Jun, Suri2013Dec, Rigetti2012Sep, Goetz2017Mar, Zhang2017Jan, Rieger2023Feb}.
For this, we follow the derivation in Refs.~\cite{Clerk2007Apr, Rigetti2012Sep} that connects photon shot noise-induced dephasing $\Gamma_{\bar{n}}$ and an AC Stark shift $\Delta f_\mathrm{q}$ of the qubit with the average photon population $\bar{n}$ in the resonator:
\begin{equation}
\label{eq:dephasing}
\Gamma_{\bar{n}} + 2\pi i\Delta f_\mathrm{q}= \frac{\kappa}{2} \left( \sqrt{\left( 1 + \frac{i \chi}{\kappa}
\right)^2 + \frac{4i \chi}{\kappa}\bar{n}} - 1 \right).
\end{equation}
The frequency shift $\Delta f_\mathrm{q}=\Delta f_{\mathrm{q},\bar{n}}+\Delta f_{\mathrm{q},0}$ consists of a photon number-dependent term $\Delta f_{\mathrm{q},\bar{n}}$ and the Lamb shift $\Delta f_{\mathrm{q},0} = (\chi / 2\pi) / 2$.
In the regime of $\vert\chi\vert\lessapprox\kappa$, we can write in the limit of small photon numbers $\bar{n}\lesssim 0.1$:
\begin{equation}
\Gamma_{\bar{n}} + 2\pi i\Delta f_{\mathrm{q},\bar{n}}=\frac{\kappa\chi}{\kappa^2+\chi^2}(\chi + i \kappa)\bar{n}.
\label{eq:eq2}
\end{equation}

In order to assess the thermalization of the flexible stripline, we implement the pulse sequence illustrated in \figrefadd{fig:meas1}{(c)}.
Before each repetition, we couple a DC pulse into the input line with a combiner to actively heat the attenuators.
This heat pulse has an amplitude of $0.5\,$V outside the cryostat, corresponding to a heat input of $0.4\,$mW on the dilution stage attenuators and orders of magnitude larger in amplitude than what is used for the readout pulse.
We then wait a variable time $t_\mathrm{cool}$ before performing a Ramsey sequence, from which we can infer the excess photon shot noise and the temperature of the attenuators.
After each repetition, we wait $2\,$ms to prevent cumulative heating.
To extract $\Gamma_{\bar{n}}$, we subtract from all measured decoherence rates $\Gamma_2^*$ the same offset value such that, for large $t_\mathrm{cool}$, $\Gamma_{\bar{n}}$ corresponds to the average value found in \figrefadd{fig:meas1}{(a)}.
Extracted values for $\Gamma_{\bar{n}}$ and $\Delta f_{\mathrm{q},\bar{n}}$ as a function of $t_\mathrm{cool}$ are depicted in \figrefadd{fig:meas1}{(d)-(e)} for three heat pulse durations $t_\mathrm{heat}\in[0.5,\,5,\,50]\,$\si{\micro\second}.

We model the data using \eqref{eq:dephasing}, assuming the input line is a single black-body radiator with Bose-Einstein distribution $\bar{n}(T)$.
The emitter's temperature rises from its thermal equilibrium $T_0$ by $\Delta T$ and relaxes exponentially with a time constant $\tau_\mathrm{cool}$ after the heat pulse:
\begin{equation}
\label{eq:heat-fit}
\begin{aligned}
\bar{n}(t_\mathrm{cool}) &= \left(\exp\left(\frac{h f_\mathrm{r}}{k_\mathrm{B}T(t_\mathrm{cool})}\right)-1\right)^{-1} \quad\mathrm{with}\\ T(t_\mathrm{cool}) &= T_0 + \Delta T \exp\left(-\frac{t_\mathrm{cool}}{\tau_\mathrm{cool}}\right),
\end{aligned}
\end{equation}
where $k_\mathrm{B}$ and $h$ are the Boltzmann and the Planck constants, respectively.

The single black body model, dominated by the temperature of the coldest attenuator, is justified by the following considerations.
All attenuators in both our setups are thermalized to temperatures $T<10\,$K, in which case the heat capacity and conduction in metals is dominated by electrons and it is therefore proportional to $T$~\cite{Pobell}.
Since the attenuation value between cryostat stages is chosen to exceed the temperature ratios (e.g. 20$\,$dB between the 4$\,$K and 200$\,$mK stage), the temperature of all attenuators heats up by a similar amount.
For the attenuator at the lowest temperature stage, this implies the largest relative temperature change.
Moreover, the hot electron effect~\cite{Roukes1985Jul, Wellstood1994Mar} and increased Kapitza phonon-phonon boundary resistance~\cite{Little1959} play a significant role at temperatures below a few hundred millikelvin, increasing even further the temperature of the last attenuator~\cite{Yeh2017Jun}. 

After a heating pulse, the thermalization of the flexible stripline temperature is modeled by a simultaneous fit of the measured relaxation curves $\Gamma_{\bar{n}}$ and $\Delta f_{\mathrm{q},\bar{n}}$ to \eqref{eq:dephasing} and \eqref{eq:heat-fit}, yielding a common time constant $\tau_\mathrm{cool}=0.28\,$ms for all three values of $t_\mathrm{heat}$~(cf. \figrefadd{fig:meas1}{(d)~-~(e)}).
In \eqref{eq:heat-fit}, we fix the effective temperature $T_0=\nolinebreak58\,$mK of the black body, corresponding to the mean residual thermal photon population in \figrefadd{fig:meas1}{(a)}.
The fit yields temperature differences $\Delta T =$\nolinebreak $[24,\,55,\,114]\,$mK.
As shown in \suppref{sec:supp:AttHeatCoax}, we repeated the same experiment for the coaxial cable setup, resulting in a time constant of $\tau_\mathrm{cool}=0.55\,$ms.
This indicates that the flexible stripline integrated attenuators are at least as well thermalized as the coaxial cable attenuators.

Further comparative measurements with the two ipnut line setups show no significant difference in qubit performance (\figref{fig:meas2}).
This includes an extraction of the qubit temperature $T_\mathrm{q}$ from $IQ$ distributions over a range of qubit frequencies $f_\mathrm{q}\in[0.285, 1.23]\,$GHz, obtained by sweeping $\Phi_\mathrm{ext}$ (\figrefadd{fig:meas2}{(a)}).
We find $T_\mathrm{q}$ for both setups to be almost constant over the whole range and close to the $\approx20\,$mK temperature of the dilution stage; $(26.4\pm2.0)\,$mK for the flexline setup, comparable to $(30.4\pm1.4)\,$mK for the coaxial cable setup.
The minor difference in $\bar{T}_\mathrm{q}$ is in the range of commonly observed fluctuations between cooldowns.
The uptake in $T_\mathrm{q}$ for both setups as the qubit frequency decreases below $0.4\,$GHz or increases beyond $1\,$GHz could be explained by the fact that the qubit population either approaches 50\% or zero, respectively. In both cases, the extracted temperature becomes susceptible to rare out-of-equilibrium excitations, for example from ionizing radiation~\cite{Vepsalainen2020Aug} or readout quantum-demolition effects~\cite{Gusenkova2021Jun}.
In literature, typical values for $T_\mathrm{q}$ fall in a broad range between 20$\,$mK and 60$\,$mK~\cite{Gusenkova2021Jun, Rieger2023Feb, Spiecker2023Sep, Vool2014Dec, Najera-Santos2024Jan, Jin2015Jun}.
Finally, \figrefadd{fig:meas2}{(b)} shows that the qubit’s energy relaxation and decoherence rates near $\Phi_\mathrm{ext}/\Phi_0 = 0.5$ remain unchanged between the two setups within our measurement accuracy.

\section{Conclusions}
In conclusion, when using a flexible stripline assembly to connect a qubit readout input line from room temperature to the dilution stage of a cryostat, we observe a residual population of the readout resonator below $3.5\cdot10^{-3}$ photons, a $0.28\,$ms thermalization time of the flexible stripline attenuators, and an effective qubit temperature of $26.4\,$mK, close to the temperature of the dilution stage.
Furthermore, we observe no significant difference in qubit performance when using flexible striplines or conventional coaxial cables. The heating pulse methodology presented here can serve as a simple health check for other groups to test the thermalization of their input lines.
These results encourage the use of flexible striplines in future cryogenic microwave setups, enabling at least an order of magnitude increase in the density of microwave input circuitry, paving the way for increasingly complex superconducting detectors and quantum devices.

\section*{Data Availability}
All relevant data are available from the corresponding author upon reasonable request.

\section*{Acknowledgements}
We are grateful to L. Radtke and S. Diewald for technical assistance. We thank J. de Groot for support with design and machining and S. Bosman for constructive feedback. We acknowledge funding from the European Commission (FET-Open AVaQus GA~899561). P.P., M.S. and N.G. acknowledge partial funding from the German Ministry of Education and Research (BMBF) within the project GEQCOS (FKZ:~13N15683). V.A. and W.W. acknowledge support from the European Research Council advanced grant MoQuOS (No.~741276). Facilities use was supported by the KIT Nanostructure Service Laboratory (NSL). We acknowledge qKit for providing a convenient measurement software framework.

\balancecolsandclearpage

\onecolumngrid
\section*{Appendices}
\vspace{0.6cm}
\twocolumngrid
\appendix

\section{Thermalization Experiments}
\label{sec:supp:thermalizationexperiments}
Before the installation of the flexible microwave stripline (\suppref{sec:supp:readoutlines_flex}), we investigated the quality of different thermal contacts.
\subsection*{Experimental setup}
\begin{figure*}[tb]
\centering
\includegraphics[width=\textwidth]{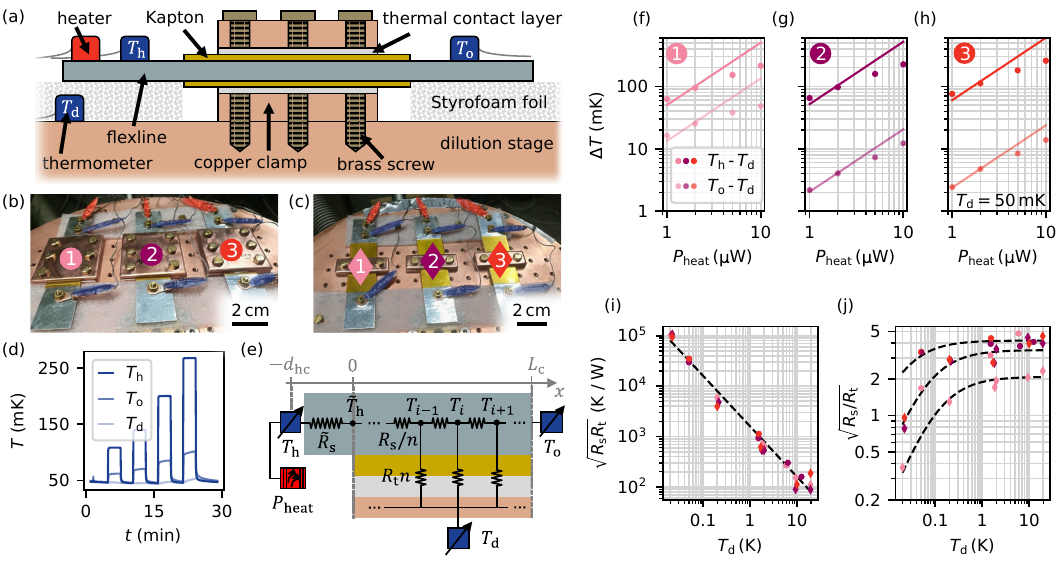}
\caption{Thermalization of flexible stripline samples.
(a)~Side-view sketch of the experiment: The central section of a flexible stripline is clamped inbetween two copper plates and attached to the dilution stage of the cryostat via brass screws. Between the copper clamps and the flexline is the factory-made \textit{Kapton$\circledR$} isolation and possibly an additional thermal contact layer. We attach a heater close to one end of the stripline and three thermometers on the heater side ($T_\mathrm{h}$), opposite to the heater side ($T_\mathrm{o}$) as well as on the dilution stage ($T_\mathrm{d}$). In order to minimize the thermal contact between the thermometers and the dilution stage, we place a Styrofoam foil between the ends and the dilution stage.
The physical implementation of this experiment is shown in
(b)~with a larger contact area (circle markers) and in
(c)~with a smaller contact area (diamond markers) between each side of the strip and the clamps. For each clamp, we investigated three different thermal contact layers: 1. no additional layer (pink), 2. grease (magenta) and 3. SPM (red). Thermometers (highlighted in blue) and heaters (in red) were fixed to the test cable with brass nuts, brass washers and a brass screw passing through the cable. Thermometers measuring $T_\mathrm{d}$ are not visible here.
(d)~Example of monitored temperatures while injecting sequentially increasing powers $P_\mathrm{heat}$ into the heater.
(e)~Simplified lumped-element model of the setup. We introduce a horizontal $x$-axis along the stripline with origin at the boundary between the heater side and the clamp.
(f)-(h) Temperature differences $T_\mathrm{h} - T_\mathrm{d}$ and $T_\mathrm{o} - T_\mathrm{d}$ for the three different thermal contacts and the large clamp (circle markers). Note that both the x- and y-axes have a logarithmic scale. The two lines are linear fits crossing the origin to the two datasets. Since we observe non-linear effects above a certain power (here approximately \qty{3}{\micro\watt}), we only fit the points below this threshold. 
(i) $\sqrt{R_\mathrm{s}R_\mathrm{t}}$ and (j) $\sqrt{R_\mathrm{s}/R_\mathrm{t}}$ for all setups and clamp sizes as a function of the dilution stage temperature $T_\mathrm{d}$, extracted via  \eqref{eq:temp_ratio} and \eqref{eq:therm_solution}. The dashed line in (i) is a $\propto T_\mathrm{d}^{-1}$-fit to all data points with a proportionality constant of \qty{1600}{\kelvin^2\per\watt}. The dashed lines in (j) are guides to the eye for three different subsets of data points, as explained in the text.
}
\label{fig:supp:clamp}
\end{figure*}
We thermalize test stripes of a flexline in their central part with a copper clamp and possibly an additional thermal contact layer to the dilution stage and then apply heat to one side of the cable (\figrefadd{fig:supp:clamp}{(a)}). 
We measure the temperatures on the heater side ($T_\mathrm{h}$), opposite to the heater side ($T_\mathrm{o}$) as well as on the dilution stage ($T_\mathrm{d}$). 
In total, we investigate three different thermal contact layers:
\begin{enumerate}
    \item no thermal contact layer
    \item \textit{Apiezon$\circledR$ N} grease
    \item silver-polymer mixture (SPM)
\end{enumerate}
In distinct cooldowns, comprising stabilized dilution stage temperatures $T_\mathrm{d}$ from 20$\,$mK up to 20$\,$K, we used two different clamp sizes: firstly, as shown in \figrefadd{fig:supp:clamp}{(b)}, a longer clamp with a length of $L_\mathrm{c}^\mathrm{(long)}~=~45\,$mm and a minimum distance between thermometer and the clamp of $d_\mathrm{hc}^\mathrm{(long)}~=~25\,$mm, and secondly, as shown in \figrefadd{fig:supp:clamp}{(c)}, a shorter clamp with $L_\mathrm{c}^\mathrm{(short)}~=~15\,$mm and  $d_\mathrm{hc}^\mathrm{(short)}~=~40\,$mm. 
The width of the stripline is always $W=22\,$mm.
An example for the monitored temperatures for setup 1 with a larger clamp is depicted in \figrefadd{fig:supp:clamp}{(d)}, where we stabilized the dilution stage at approximately $50\,$mK and heated the stripline subsequently with \qtylist{1;2;5;10}{\micro\watt}.

\subsection*{Theoretical model}
In order to obtain a quantitative measure for the thermalization of the flexline, we thermally model the setup as shown in \figrefadd{fig:supp:clamp}{(e)}: The clamped section of the stripline is assumed to have a total thermal resistance $R_\mathrm{s}$ along the stripline and $R_\mathrm{t}$ via the thermal contact layer to the clamp. 
When we discretize the contact horizontally into $n=L_\mathrm{c} / \Delta x$ sections of equal size, then each site along the strip is connected via a smaller contact resistance $R_\mathrm{s} / n$ to its neighbors and a larger contact resistance $R_\mathrm{t} \cdot n$ to the clamp. 
The copper clamp is assumed to have a constant temperature $T=T_\mathrm{d}$ over the whole length because of its significantly larger heat conductivity compared to the thermal contact layer and its good contact to the dilution stage. 
For simplicity, we neglect temperature dependencies of all thermal conductivities due to temperature gradients along the stripline. In thermal equilibrium, applying Fourier's law to the $i$-th site leads to the following heat flow balance:
\begin{equation}
\label{eq:fourier_discrete}
    \frac{T_{i-1}-T_i}{R_\mathrm{s} / n} = \frac{T_{i}-T_{i+1}}{R_\mathrm{s} / n} + \frac{T_i-T_\mathrm{d}}{R_\mathrm{t} \cdot n}
\end{equation}
In the continuous limit ($n\rightarrow\infty$), this equation becomes
\begin{equation}
\label{eq:fourier_continuous}
    T''(x) = \left(T(x) - T_\mathrm{d}\right) \frac{R_\mathrm{s}}{R_\mathrm{t}L_\mathrm{c}^2}
\end{equation}
along the clamp ($0 \leq x \leq L_\mathrm{c}$). With the boundary conditions arising from the differential version of Fourier's law, $T'(L_\mathrm{c}) = 0$ and $T'(0)=-R_\mathrm{s}P_\mathrm{heat} / L_\mathrm{c}$, the solution of the differential equation \eqref{eq:fourier_continuous} becomes:
\begin{equation}
\label{eq:Tx_solution}
    \frac{T(x) - T_\mathrm{d}}{P_\mathrm{heat}} = \sqrt{R_\mathrm{s}R_\mathrm{t}} \frac{\cosh\left( 
\sqrt{R_\mathrm{s} / R_\mathrm{t}}\left( 1 - \frac{x}{L_\mathrm{c}} \right) \right)}{\sinh\left( \sqrt{R_\mathrm{s} / R_\mathrm{t}} \right)}
\end{equation}
The temperatures on both ends of the clamp are $T(L_\mathrm{c})=T_\mathrm{o}$ and $T(0) = \tilde{T}_\mathrm{h} = T_\mathrm{h} - \tilde{R}_\mathrm{s}P_\mathrm{heat}$, with $\tilde{R}_\mathrm{s} = (d_\mathrm{hc} / L_\mathrm{c})R_\mathrm{s}$, taking the temperature gradient along the stripline at the heater-side into account. 
Solving for the measured differences results in:
\begin{equation}
\label{eq:therm_solution}
\begin{aligned}
    \frac{T_\mathrm{o} - T_\mathrm{d}}{P_\mathrm{heat}} &= \frac{\sqrt{R_\mathrm{s}R_\mathrm{t}}}{\sinh\left( \sqrt{R_\mathrm{s} / R_\mathrm{t}} \right)} \\
    \frac{T_\mathrm{h} - T_\mathrm{d}}{P_\mathrm{heat}} &= \frac{\sqrt{R_\mathrm{s}R_\mathrm{t}}}{\tanh\left( \sqrt{R_\mathrm{s} / R_\mathrm{t}} \right)}  + \frac{d_\mathrm{hc}}{L_\mathrm{c}} R_\mathrm{s}
\end{aligned}
\end{equation}
The ratio of these two equations results in a monotonically increasing function $f = f\left( \sqrt{R_\mathrm{s} / R_\mathrm{t}} \right)$:
\begin{equation}
\label{eq:temp_ratio}
    \frac{T_\mathrm{h} - T_\mathrm{d}}{T_\mathrm{o} - T_\mathrm{d}} = \underbrace{\cosh\left( \sqrt{\frac{R_\mathrm{s}}{R_\mathrm{t}}} \right) + \frac{d_\mathrm{hc}}{L_\mathrm{c}} \sqrt{\frac{R_\mathrm{s}}{R_\mathrm{t}}} \sinh\left( \sqrt{\frac{R_\mathrm{s}}{R_\mathrm{t}}} \right)}_{f = f\left( \sqrt{R_\mathrm{s} / R_\mathrm{t}} \right)},
\end{equation}
from which we can extract $\sqrt{R_\mathrm{s} / R_\mathrm{t}}$, a measure for the thermalization of the stripline. Re-insertion of the extracted $\sqrt{R_\mathrm{s} / R_\mathrm{t}}$ in \eqref{eq:therm_solution} then also yields the geometric mean $\sqrt{R_\mathrm{s} R_\mathrm{t}}$ of the thermal resistances.

\subsection*{Analysis and results}
We choose from the monitored temperatures the points in time when a thermal equilibrium along the stripe is reached. 
Since we observe a deviation from a linear dependence of the temperature differences on $P_\mathrm{heat}$, most likely due to a temperature dependence of $R_\mathrm{s,t}$, we perform linear fits crossing the origin only to the points at smaller powers (\figrefadd{fig:supp:clamp}{(f)-(h)}). 

An overview of all extracted values for the parameters $\sqrt{R_\mathrm{s} / R_\mathrm{t}}$ and $\sqrt{R_\mathrm{s} R_\mathrm{t}}$ resulting from the above equations \eqref{eq:therm_solution} and \eqref{eq:temp_ratio} is given in \figrefadd{fig:supp:clamp}{(i)-(j)} for all setups, clamp sizes and dilution stage temperatures. 
We observe that $\sqrt{R_\mathrm{s} R_\mathrm{t}}$, the geometric mean of the thermal resistances, is nearly independent of setup and clamp size and all values fall onto the same $T_\mathrm{d}^{-1}$-fit with proportionality constant \qty{1600}{\kelvin^2\per\watt}. 
In contrast, for $\sqrt{R_\mathrm{s} / R_\mathrm{t}}$, we can identify three categories of points: The weakest thermalization is indeed reached without thermal contact layer and with the smaller clamp. 
A factor of 2 improvement in $\sqrt{R_\mathrm{s} / R_\mathrm{t}}$ is obtained either with the larger clamp but no additional thermal contact layer, or with the smaller clamp and an additional layer. 
Lastly, we reach another factor of 2 improvement in $\sqrt{R_\mathrm{s} / R_\mathrm{t}}$ with the large clamp and an additional thermal contact layer. 
In general, we do not observe a significant difference between the grease and the SPM thermal contact layer.

In conclusion, for temperatures $T_\mathrm{d} \leq 50\,$mK, we observe a clear improvement of using the larger clamp and adding either grease or SPM as a thermal contact layer. 
At higher temperatures $T_\mathrm{d} \geq 1\,$K, the difference between all setups fades out within the accuracy of our experiment except for the small clamp with no additional thermal contact layer, which still shows a weaker thermalization. 
For all setups, we observe a degrading thermalization with lower temperatures.

\section{Details on readout input lines}
\label{sec:supp:readoutlines}
\subsection{Flexible stripline}
\label{sec:supp:readoutlines_flex}
\begin{figure}[tb]
\centering
\includegraphics[width=0.87\columnwidth]{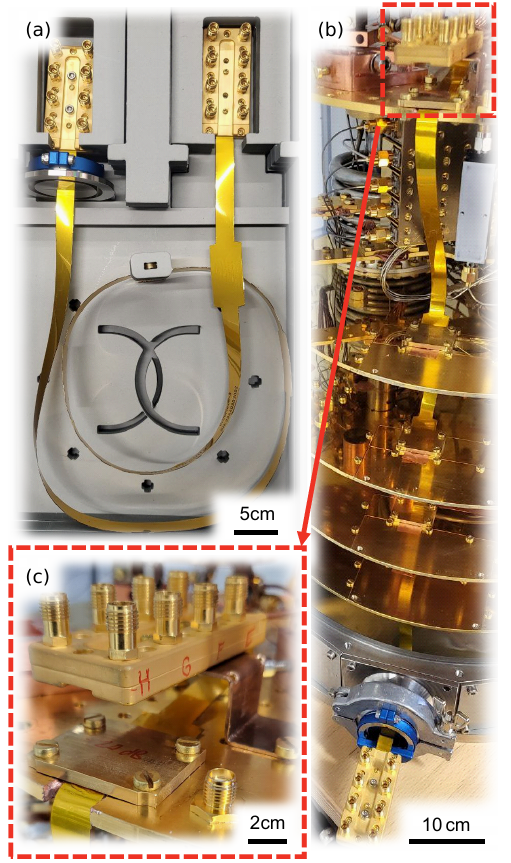}
\caption{Photographs of the flexible stripline installation. 
(a)~Assembly as delivered from \textit{Delft Circuits}. 
(b)~Fully-installed flexline (setup B in \figrefadd{fig:sample}{(b)}). 
(c)~Zoom-in of the flexible stripline thermalized at the dilution stage (red box). The thermal clamp is directly at the position of the integrated $20\,$dB attenuator. 
}
\label{fig:supp:ROLines1}
\end{figure}
The results of \suppref{sec:supp:thermalizationexperiments} show that the design of the clamp and the use of an additional thermal contact layer play a role when thermalizing the flexible stripline, especially at mK temperatures.
We clamped the flexible stripline assembly from \textit{Delft Circuits} (\figrefadd{fig:supp:ROLines1}{(a)}) within a length of $26\,$mm and with an additional layer of \textit{Apiezon$\circledR$ N} grease on both sides at the $20\,$mK, $200\,$mK, $4\,$K, $20\,$K as well as $80\,$K temperature stages (\figrefadd{fig:supp:ROLines1}{(b)}). 
Choosing only an intermediate clamp length compared to \suppref{sec:supp:thermalizationexperiments} results from a compromise between maximizing the clamp length and a facilitated integration into the cryostat. 
At the $4\,$K, $200\,$mK and $20\,$mK stages the thermal clamp is directly at the position of the integrated $20\,$dB attenuators, which are composed of four $5\,$dB unit cells in order to reduce the noise temperature~\cite{Monarkha2024May, Yeh2017Jun}.

\subsection{Coaxial line}
\label{sec:supp:readoutlines_coax}
\begin{figure}[tb]
\centering
\includegraphics[width=0.92\columnwidth]{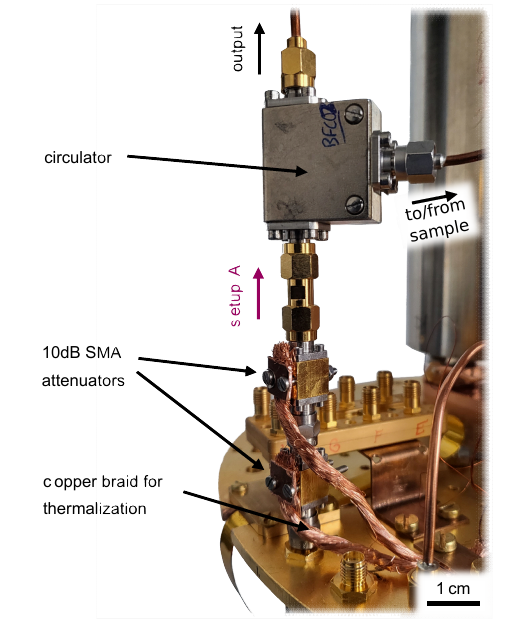}
\caption{
Overview photograph of the dilution stage part of the coaxial input line (setup A in \figrefadd{fig:sample}{(b)}). We clamp copper braids to the outer casing of the two $10\,$dB attenuators to improve the thermal contact between them and the dilution stage. Input signals are routed first to reflect on the sample and then towards the output line, using a \textit{Quinstar} circulator with $20\,$dB in-band ($4-8\,$GHz) isolation.
}
\label{fig:supp:ROLines2}
\end{figure}
We compare the experiments performed with the flexible stripline setup with one of our conventional coaxial cable setups. 
For this, we distributed the SMA connectorized attenuators on the different temperature stages in the same way as the integrated attenuators of the flexible stripline ($20\,$dB at $4\,$K, $200\,$mK and $20\,$mK each). 
At $4\,$K and $200\,$mK, we used attenuators with a stainless steel casing from \textit{XMA Corporation}. 
At $20\,$mK, we split the attenuation into two $10\,$dB attenuators with an oxygen-free high thermal conductivity (OFHC) copper outer casing from \textit{Quantum Microwave} (\figref{fig:supp:ROLines2}). 
Additionally, we attached copper braids from their outer casing to the dilution stage to improve the thermal contact.

\section{Extraction of resonator bandwidth and dispersive shift}
\label{sec:supp:ChiKappaPower}
\begin{figure}[tb]
\centering
\includegraphics[width=\columnwidth]{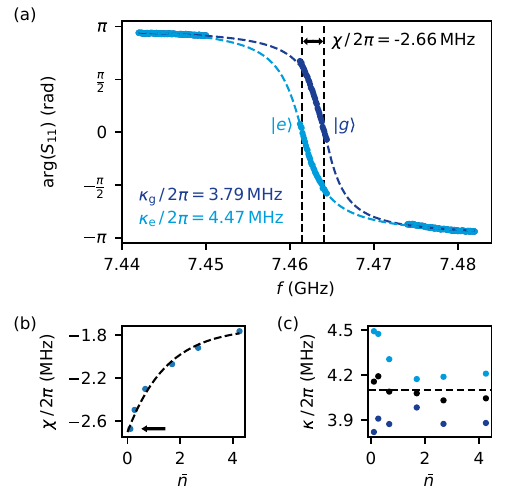}
\caption{
Qubit state dependent response of the readout resonator. (a)~Phase response of the readout resonator coupled to the qubit being in its ground state $\vert g \rangle$ (dark blue) or its excited state $\vert e \rangle$ (light blue). For each readout frequency, we measured $10^5$ consecutive reflection measurements. For each phase response associated to one of the qubit states, we perform a separate fit (dashed lines), from which we extract a dispersive shift between qubit and resonator of $\chi / 2\pi = -2.66\,$MHz and corresponding bandwidths $\kappa_\mathrm{g} / 2\pi = 3.79\,$MHz and $\kappa_\mathrm{e} / 2\pi=4.47\,$MHz. The readout strength corresponds to $\bar{n}\approx0.16$ photons on average in the resonator, as indicated by the arrow in panel (b). 
(b)~Dispersive shift $\chi$ and (c)~resonator bandwidths $\kappa_\mathrm{g} / 2\pi$ (dark blue), $\kappa_\mathrm{e} / 2\pi$ (light blue) and $\kappa / 2\pi$ (black) as a function of the average photon number $\bar{n}$ in the resonator. The black dashed line in (b) is an exponential guide to the eye. The horizontal line in (c) highlights the approximate independence of $\bar{n}$.
}
\label{fig:supp:ChiKappa}
\end{figure}
As we have seen in the main text, the extraction of the dispersive shift $\chi$ and resonator bandwidth $\kappa$ is essential for the photon number calculation (see also \eqref{eq:dephasing} and \eqref{eq:eq2}).
Since there is no additional resonator drive during the free evolution of the qubit in a Ramsey or echo experiment, we are interested in determining $\chi$ and $\kappa$ for the lowest possible photon numbers in the resonator.
For this, we measured $I$ and $Q$ quadratures, as shown in the inset plot of \figrefadd{fig:meas2}{(a)}, and extracted from these the qubit state dependent phase response of the readout resonator, as a function of the readout frequency (\figrefadd{fig:supp:ChiKappa}{(a)}).
By fitting two distinct phase responses $\mathrm{arg}(S_{11})$ to our measured data, we are able to extract $\chi$ and $\kappa$.
We repeat this procedure for different readout strengths to estimate $\kappa$ and $\chi$ for $\bar{n}\rightarrow0$.

For the dispersive shift, we observe a photon number dependence, similar as in~\cite{Gusenkova2021Jun}, with an extrapolated value of $\chi / 2\pi = -2.70\,$MHz for lowest photon numbers $\bar{n}\rightarrow0$ (\figrefadd{fig:supp:ChiKappa}{(b)}).
We calculate the mean bandwidth $\kappa / 2\pi = 0.5(\kappa_\mathrm{g} + \kappa_\mathrm{e})/2\pi$, which stays approximately constant $\kappa / 2\pi = 4.10\,$MHz over the whole range of $\bar{n}$ (\figrefadd{fig:supp:ChiKappa}{(b)}). 
We observe no significant difference between the flexline and the coaxial cable setup.
\hfill\\

\section{Attenuator heating experiment for coaxial cable setup}
\label{sec:supp:AttHeatCoax}
\begin{figure}[H]
\centering
\includegraphics[width=\columnwidth]{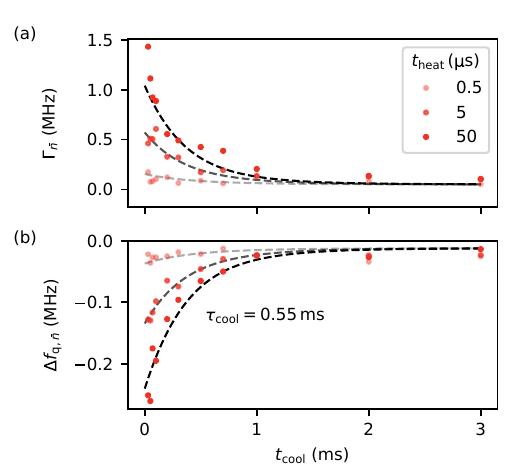}
\caption{
Attenuator heating experiment, as described in \figrefadd{fig:meas1}{(c)}, for the coaxial cable setup. 
(a)~Photon shot noise-induced dephasing rates $\Gamma_{\bar{n}}$ and 
(b)~qubit frequency shifts $\Delta f_{\mathrm{q},\bar{n}}$ extracted from Ramsey measurements. The dashed lines are a fit with common temperature relaxation time $\tau_\mathrm{cool}=0.55\,\mathrm{ms}$ for all heat pulses, following \eqref{eq:heat-fit} and \eqref{eq:dephasing}.}
\label{fig:supp:AttHeatCoax}
\end{figure}

We repeated the attenuator heating experiment from \figrefadd{fig:meas1}{(c)-(e)} with the coaxial cable setup (\figref{fig:supp:AttHeatCoax}).
Again, from interleaved energy relaxation and echo measurements over a duration of 5 hours, we extract an average photon population in the resonator of $\bar{n}=6.5\cdot10^{-3}$.
As for the flexible stripline, we calculate from this a temperature $T_0 = 71\,$mK of the black body generating the photon shot noise.
We also assume for all three heat pulses different $\Delta T$ but the same $\tau_\mathrm{cool}$ as fit parameters, resulting in temperature differences $\Delta T = [20,\,63,\,98]\,$mK and a time constant of $\tau_\mathrm{cool}=0.55\,$ms, a factor of two slower than the flexible stripline setup.
For the data with the longest heat pulse ($t_\mathrm{heat}=50\,$~\si{\micro\second}), a model based on two black bodies in the input line decaying with different time constants could possibly describe the data better.
\end{document}